\documentclass[%
 reprint,
nofootinbib,
 amsmath,amssymb,
 aps,
]{revtex4-2}

\usepackage{textcomp}
\usepackage{amsmath}
\usepackage{amsfonts}
\usepackage{amssymb}
\usepackage{graphicx}
\usepackage{hyperref}
\usepackage{subfigure}
\usepackage[usenames, dvipsnames]{color}
\usepackage[utf8]{inputenc}
\usepackage{soul}

\numberwithin{equation}{section}
\usepackage{etoolbox}
\usepackage{breqn}
\usepackage{lipsum} % just for the example
\newtheorem{theorem}{Theorem}[section]

\newtheorem{proposition}[theorem]{Proposition}

\makeatletter
\preto\maketitle{%
  \begingroup\lccode`~=`,
  \lowercase{\endgroup
  \let\saved@breqn@active@comma~% save breqn active comma
  \let~}\active@comma % set the active comma to what revtex4-1 wants
}
\appto\maketitle{%
  \begingroup\lccode`~=`,
  \lowercase{\endgroup
  \let~}\saved@breqn@active@comma % undo the change
}
\makeatother

\begin{document}
\title{Generic instabilities in the relativistic Chapman-Enskog heat conduction law}
\author{Ana L. García-Perciante${}^{1}$}\email{algarcia@correo.cua.uam.mx}
\author{Marcelo E. Rubio${}^{2}$}\email{merubio@famaf.unc.edu.ar}
\author{Oscar A. Reula${}^{3}$}\email{reula@famaf.unc.edu.ar}
\affiliation{${}^{1}$Dpto. de Matemáticas Aplicadas y Sistemas, Universidad Autónoma Metropolitana-Cuajimalpa\\
(05348) Cuajimalpa de Morelos, Ciudad de México\\
${}^{2}$Instituto de Astronomía Teórica y Experimental (CONICET)\\
Observatorio Astronómico de Córdoba, (X5000BGQ) Córdoba, Argentina\\
${}^{3}$Instituto de F\'{\i}sica Enrique Gaviola, CONICET\\
Facultad de Matem\'atica, Astronom\'ia, F\'isica y Computaci\'on\\
Ciudad Universitaria, (5000) C\'ordoba, Argentina}

\date{\today}

\begin{abstract}
We address the well-posedness of the Cauchy problem corresponding to the relativistic fluid equations, when coupled with the heat-flux constitutive relation arising within the relativistic Chapman-Enskog procedure. The resulting system of equations is shown to be non hyperbolic, by considering general perturbations over the whole set of equations written with respect to a generic time direction. The obtained eigenvalues are not purely imaginary and their real part grows without bound as the wave-number increases. Unlike Eckart's theory, this instability is not present when the time direction is aligned with the fluid's direction. However, since in general the fluid velocity is not surface-forming, the instability can only be avoided in the particular case where no rotation is present.
\end{abstract}

\maketitle

%\tableofcontents

\section{Introduction}

The Chapman-Enskog (CE) method consists on a formal expansion procedure for the distribution function of any fluid theory, under the assumption that the characteristic (macroscopic) length of the gradients of state variables is \textit{large} when compared to the (microscopic) mean free path of the molecules \cite{Chapman, Kremer, Cercignani}. This method is considered the most successful one in order to establish, in a rigorous way, the transport coefficients of a dilute (but still not rarefied) gas of classical mono-atomic molecules with no internal degrees of freedom. Moreover, it has been successfully extended to relativistic gases, even though it has been argued that the first solution-- which corresponds to the Navier-Stokes regime-- leads to nonphysical behaviour of the corresponding fluctuations in the linear approximation. However, recent numerical studies point to this method as the most appropriate one for the description of particular systems, like the ultra-relativistic electron gas \cite{GabbanaPRE17,GabbanaPRE19}.

One of the issues that arise when studying the system of equations describing transport phenomena in Relativistic Hydrodynamics is their stability, as well as the corresponding initial-value problem. In effect, a stability analysis of Eckart theory \cite{Eckart40} was carried out by Hiscock and Lindblom \cite{HL}. In that work, the authors consider linear perturbations of the equilibrium states of a relativistic fluid theory with a heat conduction law such that the hydrodynamic four-acceleration is considered a \textit{heat driving force}, namely
\begin{equation}\label{1}
q^{\mu} \sim h^{\mu\nu}\left(\frac{\nabla_{\nu}T}{T}+u^{\alpha}\nabla_{\alpha}u_{\nu}\right).
\end{equation}
Their study was carried out over a flat background with respect to $L^{2}$ perturbations, where the spatial Fourier modes were given by plane waves whose frequencies are the roots of the corresponding dispersion relation. Their main result is that such equation always leads to \textit{growing modes}, concluding that the solutions are \textit{unbounded} with very short characteristic growth times ($\tau \sim 10^{-34}$ sec. in normal conditions). Nevertheless, they examined the stability in the fluid's \textit{co-moving} frame; that is, when the time direction is aligned with the fluid's four-velocity. Since unstable modes were found along both transverse and longitudinal directions, the study of the general case was only carried out considering the Landau frame \cite{landau2013fluid}. Similar stability analyses have been previously carried out considering bulk and shear contributions for viscosity, within the Israel-Stewart formalism \cite{Denicol_2008, Pu-Shi2010}. 

Unlike Eckart theory, the CE procedure predicts a relativistic heat flux which is driven by a \textit{density gradient} instead of the four-acceleration term, that is
\begin{equation}\label{CE-heat-cond-law}
   q^{\mu} \sim h^{\mu\nu}\left(\kappa\frac{\nabla_{\nu}T}{T} - \lambda\frac{\nabla_{\nu}n}{n}\right).
\end{equation}
Moreover, it has been recently shown that the generic instability found in \cite{HL} is actually avoided when considering perturbations in the fluid's co-moving frame \cite{PA09,PRE09}. This result seems to suggest that CE theory may give rise to well-behaved and generically stable transport equations. 

With the aim of providing a concise proof of such a conjecture, in this paper we analyze in detail the stability and initial-value problem of the relativistic hydrodynamic equations when coupled with the CE thermal constitutive equation of the type (\ref{CE-heat-cond-law}). Since this is a first-order relativistic fluid theory, we first study whether or not it belongs to the particular class of \textit{first-order divergence-type} fluid theories (DTT), which were widely studied in the literature, and the issues of hyperbolicity and well-posedness can be addressed in a natural way \cite{GL90, geroch1995relativistic, geroch1991causal, geroch2001hyperbolic, Geroch96}. Particularly, it is well-known that first-order DTT \textit{do not} constitute a well-posed initial-value problem. Thus, if CE theory can be cast as a first-order DTT, then it constitutes automatically an \textit{ill-posed} theory. We show, however, that CE heat conduction law cannot be obtained from a general DTT, unless some considerations of equilibrium configurations are set up.

We then perform a stability analysis of linear perturbations around equilibrium solutions, where the decoupling and decay of the transverse modes as well as the damping of the longitudinal modes are obtained in a general fashion, without any approximation\footnote{In \cite{PA09}, the curl of the spatial components of the momentum balance equation was calculated in order to decouple the transverse mode. Also, in \cite{PRE09}, Mountain's approximate method is used in order to examine the longitudinal modes.}. Although the corresponding perturbations of the CE system of equations are found to be stable from the fluid's co-moving reference frame, we show that this is not the case when considering perturbations over the whole set of equations written with respect to a generic time-direction, as short wavelength fluctuations may grow arbitrarily fast. This general case turns out to be essential for the description of rotational fluids, as the four-velocity may not be surface-forming, being the instability \textit{generic} in that sense.

In order to address the points described above, this paper is organized as follows. In Section \ref{DTT-sec}, a brief discussion about CE theory as a first-order DTT is included. Section \ref{CE-sec} is dedicated to a general treatment for the perturbations of the transport equations considering the CE procedure, showing that: no instabilities are present when evolving along the fluid's four-velocity direction, but exponential growing fluctuations may appear when evolving with respect to a general time-direction, concluding the ill-posedness of the theory. Finally, a thorough discussion of our results and final remarks are included in Section \ref{discussion-sec}.

Throughout this work, we consider the $(-,+,+,+)$ signature for the spacetime background metric $g_{\mu\nu}$. Also, we adopt a system of units such that $c = m = k_B  = 1$, where $c$ is the speed of light in vacuum, $m$ is the rest mass of the molecules, and $k_B$ is Boltzmann constant.

\section{Chapman-Enskog theory and first-order DTT}
\label{DTT-sec}

With the aim to study the hyperbolicity of the relativistic fluid equations within the CE formalism, we first inspect the viability to write the theory as a first-order DTT. In this class of fluid theories, the dynamical variables are the particle-density current $N^{\mu}$ and the energy-momentum tensor $T^{\mu\nu}$, both satisfying local conservation laws. In order to take into account dissipative effects, an additional tensor field $A^{\mu\nu\sigma}$ is introduced as an algebraic function of $N^{\mu}$ and $T^{\mu\nu}$. Such a tensor satisfies the equation 
\begin{equation} \label{Aabc-equation}
\nabla_{\mu} A^{\mu\nu\sigma} = I^{\nu\sigma},
\end{equation}
where $I^{\mu\nu}$ is a source-like term, and all constitutive relations are obtained from it. There also exists a local entropy-density current $S^{\mu}$ such that, as a consequence of the dynamical equations, its divergence is non-negative. A brief review about DTT is presented in Appendix \ref{app-A}. 

These conditions are, by nature, phenomenological, and serve as a link to the microscopic approach, from which particle-flux, energy, momenta and entropy-flux can be obtained from a distribution function $f$; namely,
    \begin{eqnarray}
        N^{\mu} &=& \int p^{\mu}f\,d\Omega\, \\
        T^{\mu\nu} &=& \int p^{\mu}p^{\nu}f\,d\Omega\, \\
        S^{\mu} &=& \int p^{\mu}f\ln(f) \, d\Omega\,
    \end{eqnarray}
where the integrals are taken over the mass shell $p^{\mu}p_{\mu} = -1$. The constitutive tensor $A^{\mu\nu\sigma}$ is obtained from the third moment of $f$, by the following relation:
\begin{equation}
A^{\mu\nu\sigma}=\int \left(p^{\mu}p^{\nu}p^{\sigma}+\frac{1}{4}p^{\mu}g^{\nu\sigma}\right)f\,d\Omega\,,\label{37}
\end{equation}
where the second term which is proportional to the background metric is included in order to enforce the trace-free condition $g_{\nu\sigma}A^{\mu\nu\sigma} = 0$ (see Appendix \ref{app-A}).

An interesting consequence of all the above requirements is the existence of a local scalar \textit{generating function} $\chi$ that depends on a new set of variables $\{\xi,\xi_{\mu},\xi_{\mu\nu}\}$ such that \cite{geroch1991causal}
\begin{equation}\label{change-var1}
    N^{\mu} = \frac{\partial^2\chi}{\partial\xi\partial\xi_{\mu}}\,,\quad T^{\mu\nu} = \frac{\partial^2\chi}{\partial\xi_{\mu}\partial\xi_{\nu}}\,,
\end{equation}
and
\begin{equation}\label{change-var2}
    A^{\mu\nu\sigma} = \frac{\partial^2\chi}{\partial\xi_{\mu}\partial\xi_{\nu\sigma}} - \frac{g_{\alpha\beta}}{4}\frac{\partial^2\chi}{\partial\xi_{\mu}\partial\xi_{\alpha\beta}}g^{\nu\sigma}.
\end{equation}
The variable $\xi_{\mu}$ is associated with the fluid four-velocity degrees of freedom, while dissipative degrees of freedom are associated with those coming from $\xi_{\mu\nu}$. Moreover, the entropy production through dissipation is found to be $\sigma = - I^{\mu\nu}\xi_{\mu\nu}$. 

First-order DTT are obtained from generating functions which are \textit{linear} in $\xi_{\mu\nu}$. This implies that, by definition, the constitutive tensor $A^{\mu\nu\sigma}$ cannot depend on $\xi_{\mu\nu}$. Thus, it must be an algebraic function of only $u^{\mu}$ and the metric tensor. The most general third-rank tensor field satisfying all these requirements reads
\begin{equation}\label{tensorA-macro-zeroth}
A^{\mu\nu\sigma} = A_o u^{\mu} u^{\nu} u^{\sigma} + A_1 u^{\mu} g ^{\nu\sigma} + \left(A_o - 4A_1\right) g^{\mu(\nu}u^{\sigma)}.
\end{equation}
Given that only \textit{local equilibrium} quantities may appear in $A^{\mu\nu\sigma}$ for first-order theories\footnote{In the context of DTT, tensor $I^{\mu\nu}$ contains all dissipative fluxes, while the divergence of $A^{\mu\nu\sigma}$ has information about the driving forces. Since the latter are given in terms of the \textit{gradients} of state variables, only local equilibrium quantities may appear in $A^{\mu\nu\sigma}$ when considering first-order theories.}, and that we are interested in describing a \textit{classical} relativistic ideal gas, we consider the J\"uttner
distribution function for local equilibrium \cite{Juttner1, Juttner2, Kremer}
\begin{equation}\label{38}
f^{\left(0\right)}=\frac{n}{4\pi z K_{2}\left(\frac{1}{z}\right)}\exp{\left(\frac{p_{\alpha} u^{\alpha}}{z}\right)},
\end{equation}
where $n$ is the particle density, $z = T/mc^2$ ($T$ being the local temperature of the gas and $m$ the mass of each particle) and $K_n$ stands for the $n$-th modified Bessel function of second kind (see \cite{Kremer} for definitions and properties of $K_n$). From (\ref{tensorA-macro-zeroth}), (\ref{38}) and following the calculations given in Appendix \ref{app-A}, we get
\begin{equation}\label{a0-coef}
    A_o = n\left(1 + 6z\mathcal{G}(z)\right), \quad A_1 = z\mathcal{G}(z) + \frac{1}{4},
\end{equation}
where $\mathcal{G}(z)=\frac{K_3(1/z)}{K_2(1/z)}$.

At this order, the most general source-like tensor field that can be constructed as an algebraic function of the dynamical variables reads
\begin{equation} \label{source-gral}
    I^{\mu\nu} = I_o u^{\mu}u^{\nu} + I_1 u^{(\mu}q^{\nu)} + I_2 I_{\perp}^{\mu\nu},
\end{equation}
where $I_{\perp}^{\mu\nu}$ is the part of $I^{\mu\nu}$ which is completely orthogonal to $u^{\mu}$, and $\{I_o,I_1,I_2\}$ are chosen in such a way that $g_{\mu\nu}I^{\mu\nu}=0$ and $-\xi_{\mu\nu}I^{\mu\nu} \geq 0$. Thus, Eq. (\ref{Aabc-equation}) allows to compute the heat flux by considering the projection
\begin{equation}
    q^{\gamma} = -\frac{2}{I_1} u_{\sigma}h^{\gamma}{}_{\nu}\nabla_{\mu}A^{\mu\nu\sigma},
\end{equation}
where $h^{\mu}{}_{\nu} = \delta^{\mu}{}_{\nu} + u^{\mu}u_{\nu}$. Now, using expressions (\ref{tensorA-macro-zeroth}) and (\ref{a0-coef}), a direct but careful calculation yields
\begin{eqnarray*}\label{almost-heat-flux}
q^{\gamma} &\propto& h^{\gamma\alpha}\left[\left(z\mathcal{G}'(z) + \mathcal{G}(z)\right)\nabla_{\alpha} T+ z\mathcal{G}(z)\frac{\nabla_{\alpha} n}{n}\right]\\ 
&+& \left(1 + 5z\mathcal{G}(z)\right)\dot{u}^{\gamma},
\end{eqnarray*}
where $\dot{u}^{\gamma} := u^{\alpha}\nabla_{\alpha}u^{\gamma}$ is the fluid's four-acceleration at leading order. On the other hand, conservation of energy-momentum tensor allows to express $\dot{u}^{\gamma}$ as
\begin{equation}\label{acceleration}
\dot{u}^{\gamma}= \dot{u}_{\mbox{\scriptsize{PF}}}^{\gamma} + \dot{u}_{\mbox{\scriptsize{dis}}}^{\gamma}\,,
\end{equation}
where
\begin{equation}\label{pf-acceleration}
\dot{u}_{\mbox{\scriptsize{PF}}}^{\gamma} = -\frac{z}{\mathcal{G}\left(z\right)}h^{\gamma\alpha}\left(\frac{\nabla_{\alpha}T}{T}+\frac{\nabla_{\alpha}n}{n}\right)
\end{equation}
is the four-acceleration at \textit{zeroth} order; i.e., the one that corresponds to the perfect fluid solution; while
\begin{equation}\label{dis-contr-acceleration}
    \dot{u}_{\mbox{\scriptsize{dis}}}^{\gamma} = -\frac{\dot{q}^{\gamma} + \theta q^{\gamma} + h^{\gamma}{}_{\beta}\left(q^{\alpha}\nabla_{\alpha}u^{\beta} +\nabla_{\alpha}\pi^{\alpha\beta}\right)}{n\mathcal{G}(z)}
\end{equation}
is the corresponding dissipative contribution, where $\theta = \nabla_{\alpha}u^{\alpha}$ and $\pi^{\alpha\beta} = h^{\alpha}{}_{\mu}h^{\beta}{}_{\nu}T^{\mu\nu}$. Finally, introducing Eqs. (\ref{acceleration}), (\ref{pf-acceleration}) and (\ref{dis-contr-acceleration}) into Eq. (\ref{almost-heat-flux}), one is led to
\begin{equation}\label{heat-flux-FULL}
    q^{\gamma} = q^{\gamma}_{\;\mbox{\scriptsize{CE}}} + \delta q^{\gamma},
\end{equation}
where
\begin{equation}
   q^{\gamma}_{\;\mbox{\scriptsize{CE}}} = - z \bar{\mathcal{G}}(z) h^{\gamma\alpha}\left(\frac{\kappa}{\lambda}\frac{\nabla_{\alpha}T}{T} - \frac{\nabla_{\alpha}n}{n}\right),
\end{equation}
\begin{equation} \label{first-order-contr-hf}
    \delta q^{\gamma} = \mathcal{G}(z)\left[\mathcal{G}(z) - \bar{\mathcal{G}}(z)\right]\;\dot{u}_{\mbox{\scriptsize{dis}}}^{\gamma}\,,
\end{equation}
and the function $\bar{\mathcal{G}}(z)$ is given by
\[
\bar{\mathcal{G}}(z):=\mathcal{G}(z)-\frac{1+5z\mathcal{G}(z)}{\mathcal{G}(z)}\,.
\]

Thus, we conclude that the most general first-order DTT does not lead to the CE heat flux constitutive relation. Surprisingly however, this equation is fully recovered when the contribution coming from the four-acceleration at leading order is neglected, considering just the one coming from equilibrium configurations (for it is $\delta q^{\gamma} = 0$). This fact is quite relevant, since this is one of the hypothesis of CE construction. Indeed, in order to recover well-tested laws within this method (such as Navier-Newton and Fourier laws), a key step is to substitute the time derivatives of $n$, $T$ and $u^{\mu}$ by their corresponding expressions obtained from the lower order solution (that is, from local equilibrium). If this step is not carried out, one would be led to erroneous predictions, as well as inconsistent values for the transport coefficients when compared to experimental data \cite{degroot, GabbanaPRE17, Kremer,GRG11}.

Given that the CE theory cannot be obtained from a first-order DTT, we go further and study the stability of the theory in order to conclude the main result of this article.

\section{Stability and ill-posedness}
\label{CE-sec}

In this section we inspect the stability of the fluid system of equations, together with the constitutive relation arising from CE procedure. We first perform the calculations with respect to a frame co-moving with the fluid, recovering the stability of the corresponding linear perturbations. Then, we consider perturbations considering a general time direction, and see that the roots of the dispersion relation have positive real parts. We finally conclude the ill-posedness of the theory, as there are modes which grow arbitrarily fast in the high-frequency limit. 

\subsection{Perturbations in the co-moving frame}

As it was pointed out before, by following the CE procedure one arrives to a heat flux constitutive equation which is written in terms of the gradients of dynamical variables \cite{GRG11}, namely
\begin{equation}
q^{\mu}=-h^{\mu\nu}\left(\kappa\frac{\nabla_{\nu}T}{T}-\lambda\frac{\nabla_{\nu}n}{n}\right).\label{24}
\end{equation}
The corresponding fluctuations satisfy the equation
\begin{equation}\label{25}
\delta q^{\mu}=-h^{\mu\nu}\left(\kappa\frac{\nabla_{\nu}\delta T}{T}-\lambda\frac{\nabla_{\nu}\delta n}{n}\right).
\end{equation}

By perturbing the fluid equations around an arbitrary equilibrium state, one is led to the following set of equations:
\begin{eqnarray}
\nabla_{\mu}\delta N^{\mu} &=& 0 \label{2}\\
\nabla_{\mu}\delta T^{\mu\nu} &=& 0 \label{2.5}
\end{eqnarray}
where
\begin{eqnarray}
\delta T^{\mu\nu} &=& \delta(n\varepsilon)u^{\mu}u^{\nu}+2(n\varepsilon+p)u^{(\mu}\delta u^{\nu)} \nonumber\\
&+& \delta p h^{\mu\nu} + 2u^{(\mu}\delta q^{\nu)} \label{2.5}
\end{eqnarray}
and
\begin{equation}
\delta N^{\mu} = (\delta n) u^{\mu} + n\delta u^{\mu}, \label{5}
\end{equation}
together with Eq. (\ref{25}).
For a classical relativistic ideal gas, the thermodynamic relations $p = nT$ and
\[
\varepsilon = m\left(3z+\frac{K_{1}(1/z)}{K_{2}(1/z)}\right),
\]
hold, leading to the equation
\begin{equation}
n\varepsilon+p=nm\mathcal{G}(z).\label{6}
\end{equation}
Introducing the \textit{heat capacity}
\begin{equation}
    c_n := \left(\frac{\partial \varepsilon}{\partial T}\right)_n,\label{cn}
\end{equation}
the perturbed energy-momentum tensor now reads
\begin{eqnarray}
\delta T^{\mu\nu}&=&(nc_{n}\delta T+\varepsilon\delta n)u^{\mu}u^{\nu} + 2nm\mathcal{G}(z)u^{(\mu}\delta u^{\nu)}\nonumber\\
&+& \left(n\delta T+T\delta n\right)h^{\mu\nu}+2u^{(\mu}\delta q^{\nu)}.\label{T}
\end{eqnarray}

Thus, the set of dynamical equations for the perturbations can be written
as
\begin{equation}
M^{A}{}_B\,\delta Y^{B}=0,\label{7}
\end{equation}
with
\begin{equation}
M^A{}_B = \left(\begin{array}{cccc}0 & s & ink & 0\\
nc_{n}s & 0 & nTik & ik\\
nik & Tik & n\mathcal{G}(z)s & s\\
\frac{ik\kappa}{T} & -\frac{ik\lambda}{n} & 0 & 1
\end{array}\right)\label{30}
\end{equation}
and
\begin{equation}
\delta Y^{B} = \left\{ \delta T,\,\delta n,\,\delta u_{x},\,\delta q_{x}\right\}. \label{29}
\end{equation}
By assuming plane-wave solutions
\begin{equation}
\delta Q = \delta Q_{0}\exp\left(ikx+st\right)\label{deltaQ},
\end{equation}
we straightforwardly obtain
\begin{eqnarray*}
\delta q^{1} &=& - ik\left(\frac{\kappa}{T}\delta T-\frac{\lambda}{n}\delta n\right),\\
\delta q^{2} &=& \delta q^3 = 0. \label{27}
\end{eqnarray*}
Plugging these expressions into Eq. (\ref{2.5}), we get the conditions
\begin{equation}
    \delta u^2 = \delta u^3 = 0.
\end{equation}
The dispersion relation is $\det(M^A{}_B)=0$, or equivalently,
\begin{equation}
s^{3}+\lambda k^2 b_{1}s^{2}+k^2 b_{2}s+\lambda k^4 b_{3}=0\,, \label{31}
\end{equation}
 with
\begin{eqnarray*}
b_{1} &=& \frac{1}{n\mathcal{G}(z)}\left[\frac{1}{c_{n}}\frac{\kappa}{\lambda}\left(\frac{\mathcal{G}(z)}{z}-1\right)+1\right],\\
b_{2} &=& \frac{z}{\mathcal{G}(z)}\left(1+\frac{1}{c_{n}}\right),\\
b_{3} &=& \frac{1}{n\mathcal{G}(z)c_{n}}\left(1+\frac{\kappa}{\lambda}\right).
\end{eqnarray*}
Now, from Boltzmann's equation one can obtain the identity
\begin{equation}
\frac{\kappa}{\lambda}=\frac{\mathcal{G}\left(z\right)}{z}-1,\label{kappalambda}
\end{equation}
which directly implies
\begin{eqnarray*}
b_{1} &=& \frac{1}{n\mathcal{G}(z)}\left[c_{n}\left(\frac{\mathcal{G}(z)}{z}-1\right)^2+1\right],\\
b_{3} &=& \frac{1}{nzc_{n}}.
\end{eqnarray*}
Since all coefficients $b_n$ are positive, the Routh-Hurwitz criterion \cite{Hurwitz1895}
requires only one additional condition for stability, namely $b_{1}b_{2}>b_{3}$. In this case, we directly obtain
\begin{equation}
\frac{b_{1}b_{2}}{b_{3}}=\left(\frac{z}{\mathcal{G}(z)}\right)^{2}\left(1 + c_{n}\right)\left(1+c_n \Theta^2\right)\,,\label{32}
\end{equation}
where
\[
\Theta = \frac{1}{c_{n}}\left(\frac{\mathcal{G}(z)}{z}-1\right).
\]
Using the equilibrium properties satisfied by a classic relativistic gas \cite{Kremer}
\begin{equation}
\frac{\mathcal{G}\left(z\right)}{z}>4,\quad\frac{1}{3}<\frac{1}{c_{n}}<\frac{2}{3}\label{ineq}
\end{equation}
the last stability condition follows, and all the roots of the dispersion
relation have \textit{negative} real parts, showing that the CE constitutive equation in the co-moving frame \textit{does
not} lead to generic instabilities. We shall see in the next subsection that this is not the case when considering general time directions for the evolution.

\subsection{Perturbations in an arbitrary frame}

The procedure carried out above is not general enough for establishing the well-posedness of the Cauchy problem, since the
fluid's motion may include rotation, in which case one cannot build, even locally,
space-like hypersurfaces orthogonal to $u^{\mu}$. 
In order to address the general case, one considers a Fourier-Laplace transform in a more general frame, with a wave-like Ansatz for perturbations
\[
\delta Q = \delta Q_{0} \, \exp{\left(iK\bar{x}+S\bar{t}\right)},
\]
where the corresponding Lorentz transformation to the new coordinates reads 
\begin{equation}
k=\gamma\left(K+ivS\right),\qquad s=\gamma\left(S-ivK\right).\label{33}
\end{equation}
Introducing Eq. (\ref{33}) into Eq. (\ref{31}) we now get a quartic
complex polynomial, which can be written as
\begin{equation}
    \sum_{j=0}^{4}{\left(iS\right)^{j}K^{3-j}\left(K\tau\gamma\mu(z)\alpha_{j}+i\beta_{j}\right)}=0,\label{dispgen}
\end{equation}
where $i$ is the imaginary unit and $\alpha_j$ and $\beta_j$ are given by
\begin{eqnarray*}
\alpha_{0} &=& -n\left(b_{3}-b_{1}v^{2}\right),\\
\alpha_{1} &=& -n\left(4b_{3}-2b_{1}\left(1+v^{2}\right)\right)v, \\
\alpha_{2} &=& n\left[b_{1}\left(1+v^{2}\right)^{2}+2v^{2}\left(b_{1}-3b_{3}\right)\right],\\
\alpha_{3} &=& -n\left(4b_{3}v^{2}-2b_{1}\left(1+v^{2}\right)\right)v,\\
\alpha_{4} &=& -n\left(b_{3}v^{2}-b_{1}\right)v^{2},\\
\beta_{0} &=& \left(v^{2}-b_{2}\right)v,\\
\beta_{1} &=& 3v^{2}-b_{2}\left(1+2v^{2}\right), \\
\beta_{2} &=& \left(3-b_{2}\left(2+v^{2}\right)\right)v, \\
\beta_{3} &=& \left(1-b_{2}v^{2}\right),\\
\beta_{4} &=& 0.
\end{eqnarray*}
The coefficient $\lambda$ has been expressed in terms of a characteristic microscopic time $\tau$ as $\lambda=-n\tau\mu(z)$, and the function $\mu(z)$ includes all the temperature dependence of this transport coefficient (for further references, see \cite{Kremer,GRG11}).

By defining the parameter $\zeta=\tau/L$ (where $L$ is the characteristic scale of the system), as well as the dimensionless variables $\hat{K} := LK$ and $\hat{S} := \tau S$, it is possible to rewrite Eq. (30) as 
\begin{equation}
\sum_{j=0}^{4}{\left(i\hat{S}\right)^{j}\hat{K}^{4-j}\left(\hat{K}\zeta\gamma\mu(z)\alpha_{j}+i\beta_{j}\right)}=0.\label{eq:f}
\end{equation}

The above equation can be easily analyzed
by considering the case $K=0$; i.e., \textit{homogeneous} fluctuations. In such case, one obtains
\begin{equation}
S^3\left[\lambda \gamma v^2\left(b_3 v^2-b1\right)S+1-b_2 v^2\right]=0,
\end{equation}
which admits the real solution
\begin{equation}
S=-\frac{nz c_n}{\gamma\lambda v^{2}}\left[\frac{\frac{\mathcal{G}\left(z\right)}{z}-v^{2}\left(1+\frac{1}{c_{n}}\right)}{\frac{v^{2}\mathcal{G}\left(z\right)}{z}-c_{n}-\left(\frac{\mathcal{G}\left(z\right)}{z}-1\right)^{2}}\right].\label{34}
\end{equation}
It is straightforward to show that this solution is positive, as the term in parenthesis can be shown to be negative by using the properties given in Eq. (\ref{ineq}), together with $v < 1$.

\begin{figure}
 \centerline{\includegraphics[scale=0.5]{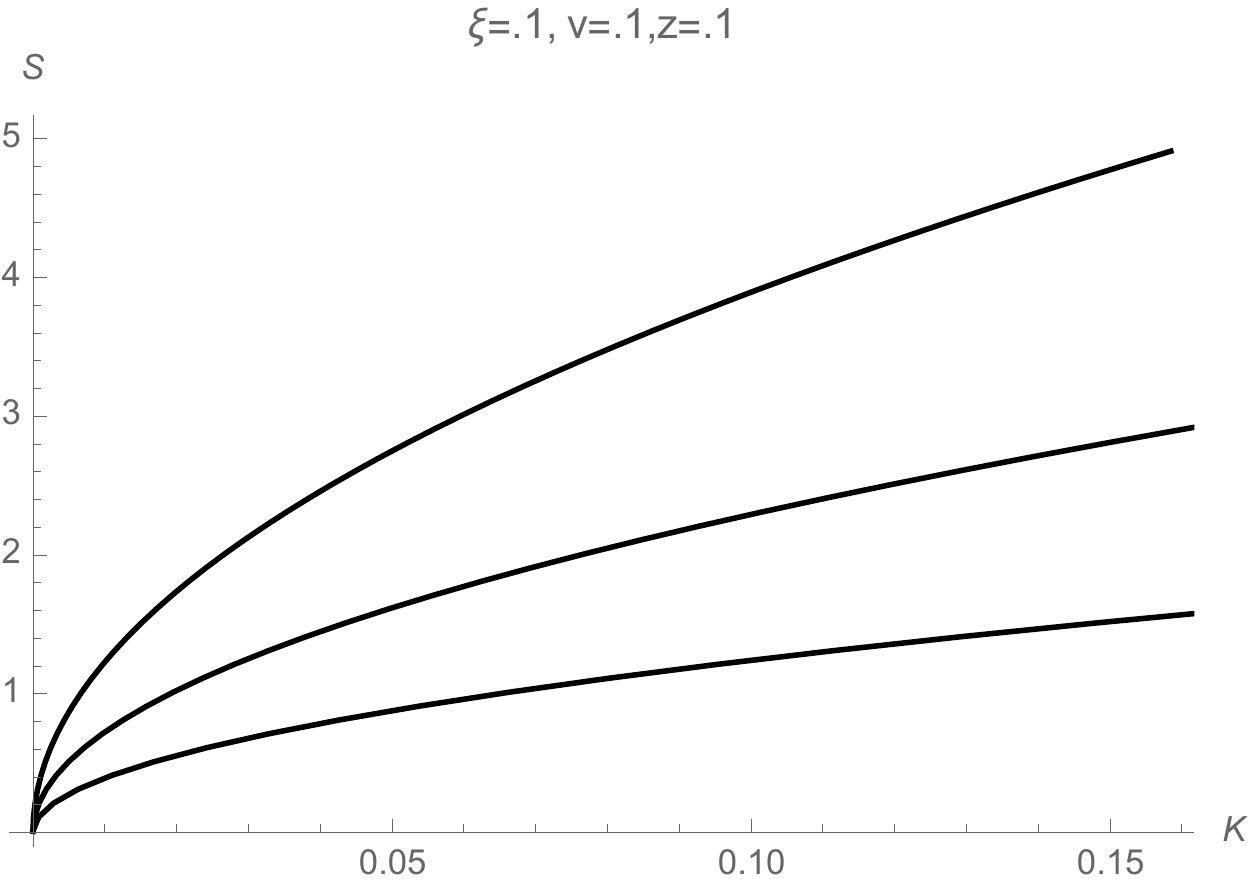}}\caption{Solutions of the dispersion relation (\ref{eq:f}) with positive real part, for $\zeta=v=z=0.1$.}\label{f1}
 \end{figure}

\begin{figure}
 \centerline{\includegraphics[scale=0.5]{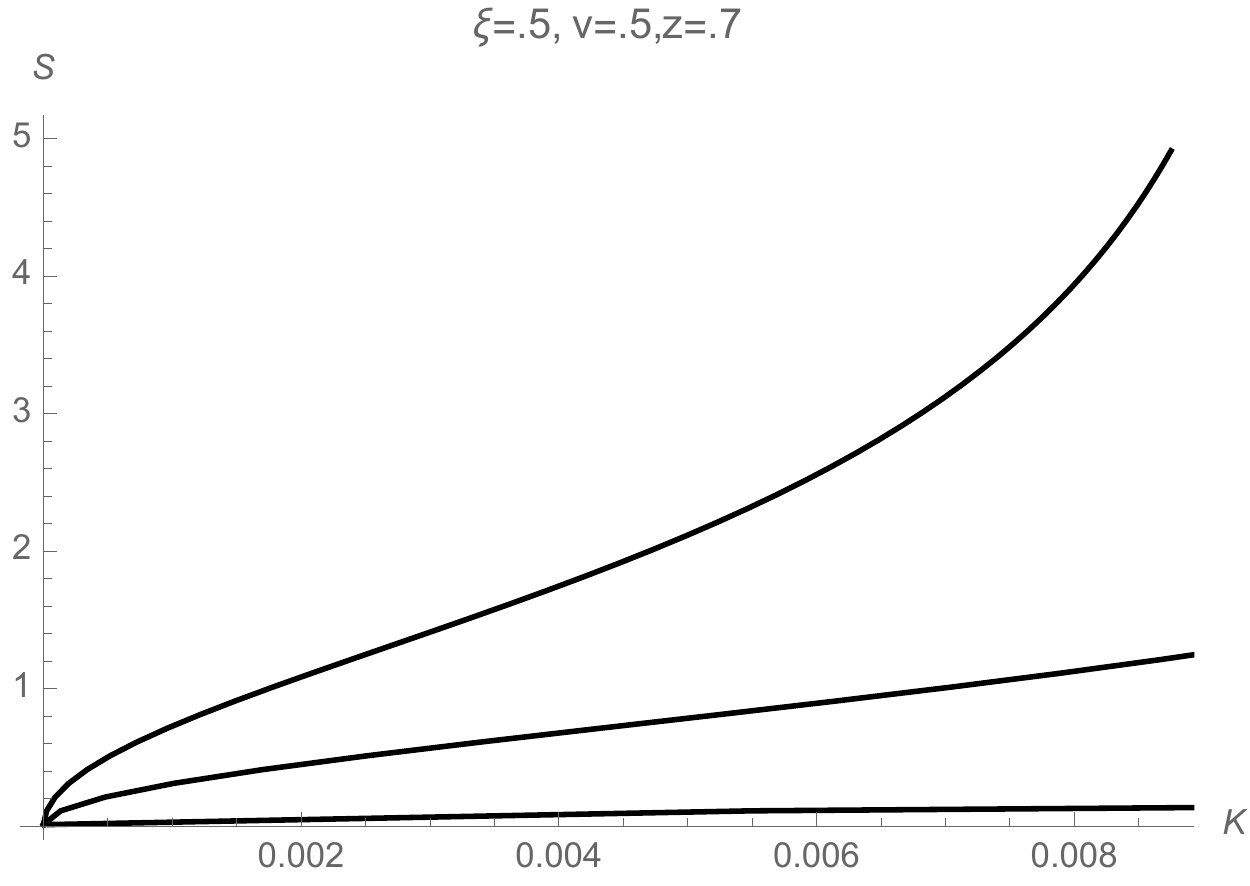}}\caption{Solutions of the dispersion relation (\ref{eq:f}) with positive real part, for $\zeta=v=0.5$ and $z=0.7$.}\label{f2}
 \end{figure}

For \textit{inhomogeneous} fluctuations ($K\neq 0$), Eq. (\ref{eq:f}) can be solved numerically for given $\zeta,\,v$ and $z$. Figures \ref{f1} and \ref{f2} show the positive real part of the 
roots of the dispersion relation as a function of $K$ for some values of the corresponding parameters.

The implication of this analysis is that, when studying the relativistic fluid equations with respect to an arbitrary time-direction (different from the co-moving one), Chapman-Enskog closure relation leads to \textit{exponentially growing} solutions and thus, the system of equations turns out to be \textit{unstable}.

\subsubsection{Ill-posedness}

As was previously introduced, the dimensionless parameter $\zeta$ is the ratio between microscopic and macroscopic scales. Since the characteristic wavelength of fluctuations is required to be \textit{large enough} for collisions to take place, it must be $\zeta\ll 1$. Then, when considering growing wave numbers $\hat{K}$, the product $\hat{K}\zeta$ should still be bounded. Taking this into account, one may look at the behaviour of those solutions $\hat{S}(\hat{K})$ of Eq. (\ref{eq:f}) which are of the form $\hat{S}=S_0+S_1 \hat{K}+S_2 \hat{K}^2$, focusing only on higher order terms (with fixed $\hat{K}\zeta$). Within this picture, it is straightforward to see that, as long as $\alpha_4\neq 0$, one has $S_2=0$. This implies that the solutions grow linearly with $\hat{K}$, i.e., $\hat{S} \sim \epsilon\hat{K}$, with $\epsilon$ satisfying
\begin{equation}
\sum_{j=0}^{4}\left(i\epsilon\right)^{j}\left(\hat{K}\zeta\gamma\mu(z)\alpha_{j}+i\beta_{j}\right)=0.\label{eq:p}
\end{equation}
Thus, we conclude that the system is not only unstable but also \textit{ill-posed}, since in the high-frequency limit ($\hat{K}\to\infty$), there are solutions that grow arbitrarily fast.

\section{Discussion and final remarks}
\label{discussion-sec}

In this article we studied  the  initial-value problem of the transport equations for a relativistic dilute gas when closed with the heat-flux constitutive equation arising from the CE formalism. Since our results are negative even in this simpler case, we have not pursued the complete case where other kinds of dissipative contributions (shear and bulk) may also be taken into account.

In order to explore the nature of the relativistic Navier-Stokes-Fourier system when the heat law is written in terms of gradients (unlike the acceleration dependence proposed by Eckart), we examined the possibility of fitting such a theory as a DTT. The purpose of such analysis lies on the fact that first order DTT have been formally shown to be ill-posed, providing thus a overwhelming argument to rule out its hyperbolicity in a direct and clean way. In this direction, we could check that the CE constitutive equation can not be obtained from Eq. (\ref{c}) in a general way, concluding that the whole system of equations do not belong to the DTT class. However, it is worth remarking that, by providing the Euler equations in order to eliminate the acceleration term in favor of space gradients of density and temperature, such an equation can be obtained. Although this substitution is consistent with Hilbert's method for the solution of the integral-differential Boltzmann equation within the CE procedure, it is inconsistent in the full framework of DTT. As a discussion, if one would like to justify such a substitution procedure, one would need to find some relation between the Knudsen parameter in the kinetic approach and the corresponding dissipative expansion within the DTT formalism. This is an open problem which is of interest and a possible direction for future work. If such justification is found to be valid, one could conclude that CE constitutive equation leads to unstable behavior at first order. In any case, it is well known that this system of equations can be made hyperbolic by adding higher order terms in dissipation, thus making it physically sound. 

The main result of this article was to show the instability of the transport equations within the CE theory with a general time-direction for the evolution. By applying an analytic criterion, we could see that this system is stable only when considering perturbations within the fluid's co-moving frame, meaning that the instability in this scheme resides not necessarily in the equations per-se, but in the flux-force relation introduced for the heat-flux. However, this result is not completely general, since the equilibrium state is assumed to be such that the hydrodynamic velocity is surface-forming. That is, since the $3+1$-decomposition is based on $u^{\nu}$, the temporal direction is fixed by it and in order for that to be the case, no rotations should be present. In order to address the stability within the CE formalism in a general scenario, an arbitrary boosted frame -with respect to the co-moving one- was considered. In this case, a positive real root was found for the corresponding dispersion relation, thus leading to exponentially growing modes. These results imply that in this general case, the system is unstable. Furthermore, the system was shown to be ill-posed, as in the high-frequency limit there are modes that grow arbitrarily fast. The underlying mathematical mechanism for the inhibition of instabilities in this particular case can be traced down to the first-order system, being parabolic in nature with either real or imaginary frequencies. A detailed analysis of this problem is presented in the pre-print \cite{Garcia-Perciante:2019}. Even though form a physical standpoint one would expect a hyperbolic system, the parabolic-damped one generated by using the CE procedure is still physically acceptable and could explain the success of such theory in numerical simulations \cite{GabbanaPRE17,GabbanaPRE19}.

\appendix

\section{}
\label{app-A}

In this appendix we give a brief review about divergence-type fluid theories, and then we compute the most general constitutive tensor for first-order DTT, considering the J\"uttner distribution function. Finally, we get an expression of the heat-flux constitutive relation predicted from this formalism by taking the appropriate projection for the divergence of the constitutive tensor.

\subsection{Brief detour on DTT}

Roughly speaking, a set of dynamic equations is considered a divergence-
type theory (DTT) if it can be written as a set of equations on the divergence of the corresponding dynamical variables. Any fluid theory which is governed
by a set of conservation laws (particle number density, energy and momentum densities, etc) constitutes, indeed, a divergence-type theory. 
However, constitutive equations, which play a major role in the assessment of causality and hyperbolicity through Geroch's
criterion need to be also expressed in divergence form. In other words, an additional tensor containing the thermodynamic forces is needed,
and the equation for its divergence should lead to the corresponding constitutive equations. 

Formally speaking, divergence-type fluid theories are the type of theories which satisfy the
following three conditions:
\begin{itemize}
    \item[\textbf{(i)}] The dynamical variables are given by the energy-momentum tensor $T^{\mu\nu}$ of the fluid and the particle number current, $N^{\mu}$;
    
    \item[\textbf{(ii)}] The dynamical equations are given by
    \begin{eqnarray}
         \nabla_{\mu}N^{\mu} &=& 0 \label{a} \\
         \nabla_{\mu}T^{\mu\nu} &=& 0 \label{b} \\
         \nabla_{\mu}A^{\mu\nu\sigma} &=& I^{\nu\sigma} \label{c}
    \end{eqnarray}
    
Here, both the constitutive tensor $A^{\mu\nu\sigma}$ and the source tensor $I^{\nu\sigma}$ are algebraic functions of the dynamical variables $N^{\mu}$ and $T^{\mu\nu}$, and $I^{\nu\sigma}$ is symmetric and traceless.

\item[\textbf{(iii)}] There exists a four-current $S^{\alpha}$, which is also a \textit{local} algebraic
function of $T^{\mu\nu}$ and $N^{\mu}$ that satisfies, \textit{as a consequence
of the dynamical equations},
\[
\nabla_{\alpha}S^{\alpha}=\sigma,
\]
with $\sigma\geq 0$ an algebraic function of $T^{\mu\nu}$
and $N^{\mu}$ and $\sigma = 0$ if and only if $I^{\nu\sigma}\equiv 0$.
\end{itemize}

In order to close the system and to have the same quantity of variables and equations, condition $g_{\mu\nu}A^{\alpha\mu\nu} = 0$ is required. Conditions (ii) and (iii) imply the existence of a generating function $\chi(\xi,\xi_{\mu},\xi_{\mu\nu})$ which contains all the information of the theory. In particular, particle flux and energy-momentum tensor can be obtained by taking derivatives of $\chi$ with respect to the corresponding variables, as it is shown in Eqs. (\ref{change-var1}) and (\ref{change-var2}). The study of hyperbolicity properties of the theory in terms of this new formulation is much more direct. In fact, by introducing a collective abstract variable $\xi^A := (\xi, \xi_{\mu},\; \xi_{\mu\nu})$, Eqs. (\ref{a}), (\ref{b}) and (\ref{c}) can be set into the form
\begin{equation}\label{geroch-form-system}
\mathcal{K}^{\mu}{}_{AB}\nabla_{\mu} \xi^B = J_A,
\end{equation}
where
\begin{equation}
\mathcal{K}^{\mu}{}_{AB} := \frac{\partial^3 \chi}{\partial \xi_{\mu} \partial \xi^A \partial \xi^B}
\end{equation}
is the \textit{principal symbol} of (\ref{geroch-form-system}) (which by construction is symmetric in the capital indices) and $J_A := (0,0,\;I_{\mu\nu})$.
Then, following Geroch's formalism \cite{Geroch96}, we say that the system is \textit{symmetric-hyperbolic} is there exists a time-like vector field $t^a$ such that the symmetric form $h_{AB} = t_a\mathcal{K}^a{}_{AB}$ is a norm; that is, it is positive-definite.

\subsection{Details of calculations of Section \ref{DTT-sec}}

We now compute the most general constitutive tensor field that can be constructed from first-order divergence-type fluid theories, using a J\"uttner distribution function in local equilibrium, $f^{(0)}$. As was pointed out in Section \ref{DTT-sec}, such a constitutive tensor is made up by means of the third moment of $f^{(0)}$.

From the macroscopic point of view, the constitutive tensor field must be an algebraic function of the dynamical variables $N^{\mu}$ and $T^{\mu\nu}$. On the other hand, since we are considering first-order theories and $A^{\mu\alpha\beta}$ includes, by definition, first derivatives with respect to the dissipative tensor, we conclude that it must be of zeroth-order (for that reason we are considering just the J\"uttner distribution function). Up to this order, both $N^{\mu}$ and $T^{\mu\nu}$ are made up in terms of the fluid four-velocity $u^{\mu}$ and the background metric $g_{\mu\nu}$. It is rather straightforward to see that the most general tensor field that satisfies these requirements, and has the symmetries imposed by the theory, is the one given in Eq. (\ref{tensorA-macro-zeroth}). In fact, recalling that the possible $p^{\mu}$ are those restricted to the mass-shell $p^{\mu}p_{\mu}=-1$ (where the mass of each fluid component is normalized to $m = 1$), we get
\begin{eqnarray}\label{a0-expl1}
 A_o &=& - \frac{2}{3} A^{\mu\alpha\beta}\left(2u_{\mu}u_{\alpha} + h_{\mu\alpha}\right)u_{\beta} \nonumber\\
 &=& 2 \int{f^{(0)}\left(-u_{\alpha}p^{\alpha}\right)\left[ \left(-u_{\alpha}p^{\alpha}\right)^2 - \frac{1}{2}\right]\,d\Omega}.
\end{eqnarray}
By the change of variables
\[
    \varepsilon = -u_{\alpha}p^{\alpha},
\]
we get 
\[
    f^{(0)} = \frac{n}{4\pi z K_2\left(1/z\right)}\,e^{-\varepsilon/z},
\]
and $d\Omega = \sqrt{\varepsilon^2 - 1}\, d\varepsilon\, dS_2$, where $dS_2$ is the area element in the unit sphere of momentum directions. Thus, integral (\ref{a0-expl1}) reduces to
\begin{eqnarray}
    A_o &=& \frac{2n}{zK_2\left(1/z\right)}\int_1^{\infty}{e^{-\varepsilon/z}\varepsilon\left(\varepsilon^2 - \frac{1}{2}\right)\sqrt{\varepsilon^2 - 1}\;d\varepsilon}\nonumber \\
    &=& \frac{2n}{zK_2\left(1/z\right)}\left(
    \frac{1}{2}\mathcal{I}_1(z) + \mathcal{I}_2(z)\right), \nonumber
\end{eqnarray}
where, for $\ell\in\mathbb{N}$,
\begin{equation}\label{I_N-integral}
    \mathcal{I}_{\ell}(z) := \int_{1}^{\infty}{e^{-\varepsilon/z}\varepsilon\left(\varepsilon^2 - 1\right)^{\ell-1/2}d\varepsilon}.
\end{equation}

The integral in Eq. (\ref{I_N-integral}) can be easily computed by means of some properties of modified Bessel functions, as shown in the following

\begin{proposition}
Let $K_{\ell}$ be the $\ell$-th modified Bessel function. Then, the following identities hold for any $\ell\in\mathbb{N}$:
\begin{enumerate}
    \item[(i)]
        \begin{equation}
            \int_{1}^{\infty}{e^{-\varepsilon/z}\varepsilon\left(\varepsilon^2 - 1\right)^{\ell-1/2}d\varepsilon} = \frac{\Gamma\left(\ell+\frac{1}{2}\right)}{\Gamma\left(\frac{1}{2}\right)}\,\left(2z\right)^{\ell}K_{\ell+1}\left(\frac{1}{z}\right).
        \end{equation}
    
    \item[(ii)]
        \begin{equation}
            \frac{dK_{\ell}(x)}{dx} = \frac{\ell K_{\ell}(x)}{x} - K_{\ell+1}(x).
        \end{equation}
\end{enumerate}
\end{proposition}

\textit{Proof.} Identity \textit{(ii)} is a direct consequence of the derivative formula
\[
    \frac{d}{dx}\left[\frac{K_{\ell}(x)}{x^{\ell}}\right] =  - \frac{K_{\ell + 1}(x)}{x^{\ell}}.
\]
Identity \textit{(i)} is also consequence of the above formula and the following important one:
\[
    K_{\ell}(x) = \left(\frac{x}{2}\right)^{\ell}\frac{\Gamma(1/2)}{\Gamma(\ell+1/2)}\int_{1}^{\infty}{e^{-xy}\left(y^2 - 1\right)^{\ell-1/2}\,dy}.
\]
$\Box$
\\

By applying the proposition above we get, then, $\mathcal{I}_1 = zK_2(1/z)$, $\mathcal{I}_2 = 3z^2 K_3(1/z)$ and
\[
A_o = n\left(1 + 6z\mathcal{G}(z)\right).
\]
Analogously, for $A_1$ we get
\begin{eqnarray}
A_1 &=& - \frac{1}{3}A^{\mu\alpha\beta}u_{\mu}u_{\alpha}u_{\beta} \nonumber \\
&=& \frac{n}{3zK_2(1/z)}\int_1^{\infty}{e^{-\varepsilon/z}\varepsilon\left(\varepsilon^2 - \frac{1}{4}\right)\sqrt{\varepsilon^2 - 1}\;d\varepsilon} \nonumber \\
&=& \frac{n}{3zK_2\left(1/z\right)}\left(\frac{3}{4}\mathcal{I}_1(z) + \mathcal{I}_2(z)\right) \nonumber \\
&=& n\left(z\mathcal{G}(z) + \frac{1}{4}\right).
\end{eqnarray}

Now, in order to obtain the Chapman-Enskog constitutive relation, we proceed to project the constitutive tensor in the space perpendicular to $u^{\mu}$. In order to do so, we find it useful to express the constitutive tensor as a sum of three contributions, namely
\begin{equation}
    A^{\mu\alpha\beta} = A_1^{\mu\alpha\beta} + A_2^{\mu\alpha\beta} + A_3^{\mu\alpha\beta},
\end{equation}
where 
\begin{eqnarray}
A_1^{\mu\alpha\beta} &=& A_o u^{\mu}u^{\alpha}u^{\beta} \nonumber\\ A_2^{\mu\alpha\beta} &=& A_1 u^{\mu}g^{\alpha\beta} \nonumber\\ A_3^{\mu\alpha\beta} &=& (A_o - 4A_1)g^{\mu(\alpha}u^{\beta)} \nonumber
\end{eqnarray}
Then,
\begin{eqnarray}
    h_{\alpha}{}^{\gamma}\nabla_{\mu}A_{1}^{\mu\alpha\beta} &=& h_{\alpha}{}^{\gamma}\left[u^{\alpha}u^{\beta}\dot{A}_o+A_o\nabla_{\mu}\left(u^{\mu}u^{\alpha}u^{\beta}\right)\right] \nonumber \\
    &=& A_o u^{\beta}\dot{u}^{\gamma},
\end{eqnarray}
yielding
\[
    u_{\beta}h_{\alpha}{}^{\gamma}\nabla_{\mu}A_{1}^{\mu\alpha\beta} = -A_o \dot{u}^{\gamma}.
\]
By similar calculations, we get
\[
    h_{\alpha}{}^{\gamma}\nabla_{\mu}A_{2}^{\mu\alpha\beta}=\left(\dot{A}_1 + A_1 \nabla_{\mu}u^{\mu}\right)h^{\gamma\beta},
\]
which implies $u_{\beta}h_{\alpha}{}^{\gamma}\nabla_{\mu}A_{2}^{\mu\alpha\beta} = 0$. Finally, defining 
\[
    A_3 := \frac{A_o}{2} - 2A_1,
\] 
we have
\begin{eqnarray}
    u_{\beta}h_{\alpha}{}^{\gamma}\nabla_{\mu}A_{3}^{\mu\alpha\beta}  &=& 2u_{\beta}h_{\alpha}{}^{\gamma}\left[g^{\mu(\alpha}u^{\beta)}\nabla_{\mu} A_3 + A_3\nabla^{(\alpha}u^{\beta)}\right] \nonumber\\
    &=& - h^{\gamma\mu}\left(\nabla_{\mu}A_3\right) + A_3\dot{u}^{\gamma} \nonumber 
\end{eqnarray}
Plugging all together, we get
\begin{eqnarray}
    q^{\gamma} &\propto& u_{\beta}h^{\gamma}{}_{\alpha}\nabla_{\mu}A^{\mu\alpha\beta} \nonumber \\
    &\propto& h^{\gamma\mu}\left(\nabla_{\mu}A_3 + A_3\frac{\nabla_{\mu}n}{n}\right) + \left(\frac{A_o}{2} + 2A_1\right)\dot{u}^{\gamma} \nonumber \\
    &=& zh^{\gamma\mu}\left(\left(z\mathcal{G}(z)\right)' \frac{\nabla_{\mu}T}{T} + \mathcal{G}(z)\frac{\nabla_{\mu}n}{n}\right) \nonumber \\ &+&\left(1+5z\mathcal{G}(z)\right)\dot{u}^{\gamma}
\end{eqnarray}
Then, we use the local expression (\ref{acceleration}) for the acceleration at the leading order, and the following formula for the derivative of $\mathcal{G}(z)$:
\begin{equation}
    \mathcal{G}'(z) = -\frac{1}{z^2}\left[\mathcal{G}^2(z)-5z\mathcal{G}(z)-1\right],
\end{equation}
which implies that
\begin{equation}
(z\mathcal{G}(z))'= \mathcal{G}(z)\left[1-\frac{\bar{\mathcal{G}}(z)}{z}\right],
\end{equation}
where
\[
\bar{\mathcal{G}}(z):=\mathcal{G}(z)-\frac{1+5z\mathcal{G}(z)}{\mathcal{G}(z)}.
\]

Finally, recalling the contribution to the heat flux $\delta q^{\gamma}$ given in Eq. (\ref{first-order-contr-hf}) by taking into account first-order corrections in the four-acceleration, we get
\begin{widetext}
\begin{eqnarray}
q^{\gamma} &\propto& h^{\gamma\mu} \mathcal{G}(z) \left[\left(1 - \frac{\bar{\mathcal{G}}(z)}{z}\right)\frac{\nabla_{\mu}T}{T} + \frac{\nabla_{\mu}n}{n}\right] - \frac{1+5z\mathcal{G}(z)}{\mathcal{G}(z)} h^{\gamma\mu}\left(\frac{\nabla_{\mu}T}{T} + \frac{\nabla_{\mu}n}{n}\right) + \delta q^{\gamma}  \nonumber \\
&=& h^{\gamma\mu} \mathcal{G}(z) \left[\left(1 - \frac{\bar{\mathcal{G}}(z)}{z}\right)\frac{\nabla_{\mu}T}{T} + \frac{\nabla_{\mu}n}{n}\right] + h^{\gamma\mu}\left[\bar{\mathcal{G}}(z) - \mathcal{G}(z)\right]\left(\frac{\nabla_{\mu}T}{T} + \frac{\nabla_{\mu}n}{n}\right) + \delta q^{\gamma} \nonumber \\
&=& h^{\gamma\mu}\bar{\mathcal{G}}(z)\left[\left(1-\frac{\mathcal{G}(z)}{z}\right)\frac{\nabla_{\mu}T}{T} + \frac{\nabla_{\mu}n}{n}\right] + \delta q^{\gamma} \nonumber \\
&=& -h^{\gamma\mu}\bar{\mathcal{G}}(z)\left(\frac{\kappa}{\lambda}\frac{\nabla_{\mu}T}{T} - \frac{\nabla_{\mu}n}{n}\right) + \delta q^{\gamma}.
\end{eqnarray}
\end{widetext}

\section*{Acknowledgments}
We would like to thank CONICET and SECyT-UNC for partial support. M.E.R is a postdoctoral fellow from CONICET, Argentina.

\bibliography{CEDTT.bib}%
\bibliographystyle{unsrt}

\end{document}